\documentclass[aps,amsmath,amssymb,twocolumn,showpacs,superscriptaddress,prb]{revtex4} 

\usepackage{graphicx}

\newcommand{\bea}{\begin{eqnarray}} 
\newcommand{\eea}{\end{eqnarray}}

\newcommand{\be}{\begin{equation}} 
\newcommand{\ee}{\end{equation}} 
 
\newcommand{\beq}{\begin{eqnarray}} 
\newcommand{\eeq}{\end{eqnarray}}

\def\H1{\widehat{H}_1}

\newcommand{\bgamma}{{\mbox{\boldmath$\gamma$}}}
 
\begin{document}
 
\title{Quantum phase diagram of the generalized ionic Hubbard model 
for AB$_{n}$ chains }

\author{M.E. Torio}
\affiliation{Instituto de F\'{$\!\!\!\!\: \iota \!\!\;$}sica Rosario, 
Consejo Nacional de Investigaciones 
Cient\'{$\!\!\!\!\; \iota \!\!\;$}ficas y
T\'{e}cnicas and Universidad Nacional de Rosario, 
Boulevard 27 de Febrero 210 Bis, 2000 Rosario, Argentina}

\author{A.A. Aligia}
\affiliation{Centro At\'{o}mico Bariloche and Instituto Balseiro, 
Comisi\'{o}n Nacional de Energ\'{$\!\!\! \iota$}a At\'omica, 8400 
Bariloche, Argentina}

\author{G.I. Japaridze}  
\affiliation{Institute of Physics, Georgian Academy of Sciences, 
Tamarashvili 6, 0177 Tbilisi, Georgia}

\author{B. Normand} 
\affiliation{D\'{e}partement de Physique, Universit\'{e} de Fribourg, 
CH-1700, Fribourg, Switzerland}
\affiliation{Centro At\'{o}mico Bariloche and Instituto Balseiro, 
Comisi\'{o}n Nacional de Energ\'{$\!\!\! \iota$}a At\'omica, 8400 
Bariloche, Argentina}
 
 
\begin{abstract} 

We investigate the ground-state phase diagram of the Hubbard model for 
the AB$_{N-1}$ chain with filling $1/N$, where $N$ is the number of atoms 
per unit cell. In the strong-coupling limit, a charge transition takes 
place from a band insulator (BI) to a correlated insulator (CI) for 
increasing on-site repulsion $U$ and positive on-site energy difference 
$\Delta$ (energy at A sites lower than at B sites). In the weak-coupling 
limit, a bosonization analysis suggests that for $N > 2$ the physics 
is qualitatively similar to the case $N = 2$ which has already been 
studied: an intermediate phase emerges, which corresponds to a bond-ordered 
ferroelectric insulator (FI) with spontaneously broken inversion symmetry. 
We have determined the quantum phase diagram for the cases $N = 3$ and 
$N = 4$ from the crossings of energy levels of appropriate excited states, 
which correspond to jumps in the charge and spin Berry phases, and from 
the change of sign of the localization parameter $z_{L}^{c}$. From these 
techniques we find that, quantitatively, the BI and FI phases are broader 
for $N > 2$ than when $N = 2$, in 
agreement with the bosonization analysis. Calculations of the Drude weight 
and $z_{L}^{c}$ indicate that the system is insulating for all parameters, 
with the possible exception of the boundary between the BI and FI phases. 

\end{abstract} 
 
\pacs{71.10.Hf, 71.45.Lr, 78.20.Bh} 
 
\maketitle 
 
\section{Introduction} 
 
The half-filled Hubbard chain with alternating on-site energies  
$\pm {\frac{1}{2}}\Delta$, known as the ionic Hubbard model (IHM),
was proposed \cite{hub,nag} to describe the neutral-ionic 
transition in mixed-stack charge-transfer organic crystals such as 
tetrathiafulvalene-$p$-chloranil.\cite{Lemee,Horiuchi} Interest in the 
model increased in the last decade due to its potential application to 
ferroelectric perovskites.\cite{Egami,res,ort,res2,gid,fab,fab2,tor} 
Although some details of the phase diagram, excitations, and expected 
physical properties of certain phases remain to be established, 
recent research has revealed the essential physics of the 
model.\cite{fab,fab2,tor,zha,brune,ali,man,dimer,poldi,otsu} 

It is self-evident that in the strong-coupling limit (hopping $t 
\rightarrow 0$), the ground state is a band insulator (BI) for on-site 
repulsion $U < \Delta$, when the sites with lower diagonal energy are 
doubly occupied, but is a type of Mott insulator (MI) for $U > \Delta$, 
when all sites are singly occupied. However, for finite, small $t$, 
perturbational approaches become invalid at $U = \Delta$ and 
non-trivial charge fluctuations persist even in the strongly coupled 
limit. Effective models around this limit have been proposed and 
analyzed\cite{poldi} but not yet studied numerically. Nevertheless, 
following the initial proposal of Fabrizio {\it et al}.,\cite{fab} 
subsequent numerical studies,\cite{tor,zha,man,otsu} and the obtaining 
of exact results for a closely related model,\cite{dimer,poldi} it has 
become clear that the IHM\ chain has two transitions as $U$ is increased. 
The first is a charge transition at $U = U_{c}$, thought to be of the Ising 
type,\cite {fab,fab2,otsu} from the BI to a bond-ordered, spontaneously 
dimerized, ferroelectric insulator (FI). The second, when $U$ is further 
increased, involves a vanishing of the spin gap at $U_{s} > U_{c}$ at 
a Kosterlitz-Thouless transition between the FI and the MI.\cite{fab,fab2} 
The phase diagram has been constructed in full detail by following the 
crossing of appropriate excited energy levels, which for this model turns 
out to be equivalent to the method of jumps in Berry phases (topological 
transitions).\cite{tor} For finite chains it has been shown that the 
topological transition at $U_{c}$ may be detectable in measurements of 
transport through annular molecules or nanodevices.\cite{dots} 
 
A particularly interesting feature of the model is the ferroelectric 
nature of the intermediate FI phase.\cite{dimer,poldi,wil} This phase 
results from an electronically induced Peierls instability, which 
generates a ferroelectric state with no ionic displacement. Interestingly, 
the elementary excitations of the FI phase have a fractional charge, 
which is proportional to the polarization.\cite{poldi} Experimentally, 
a bond-ordered ferroelectric state has been observed in the 
pressure-temperature phase diagram of the prototypical compound 
tetrathiafulvalene-$p$-chloranil.\cite{Lemee,Horiuchi} As noted by 
Egami {\it et al.},\cite{Egami} the microscopic origin of the 
displacive-type ferroelectric transition in covalent perovskite 
oxides such as BaTiO$_{3}$ remains unclear. 

The motivation for the present study is to investigate the changes in 
this type of phase diagram with the periodicity of the lattice, while 
retaining one electronegative ion and two electrons per unit cell, 
in such a way that for $U = 0$ the system is clearly a BI. However, 
away from half-filling, {\it i.e.}~for $N > 2$, the conventional Umklapp 
scattering term, which in the half-filled Hubbard chain is responsible for 
dynamical generation of a charge gap, is absent, and therefore one might 
expect metallic behavior or at minimum some qualitatively different 
physical properties at larger values of $U$ when $N \neq 2$. The 
model with $N = 3$ may also be relevant for the properties of doped, 
halogen-bridged binuclear metal chains (referred to as $MMX$ chains) 
such as R$_4$[Pt$_2$(P$_2$O$_5$H$_2$)$_4$X.$n$H$_2$O.\cite{yama} 
These compounds are related in turn to the quasi-one-dimensional 
$MX$ complexes, which have been of interest for several decades in 
the broader context of the physics of electronic chain systems. 

The paper is organized as follows. In Sec.~II we study the limit  
$t \rightarrow 0$ to derive some results for the charge and spin gaps, 
and for the effective spin Hamiltonian. In Sec.~III the opposite limit, 
of weak interactions, is treated by bosonization, and qualitative results 
are presented for the expected phases and correlation functions. 
Section IV contains a description of the numerical tools which are 
used in Sec.~V to deduce the quantum phase diagram and to discuss the 
metallic or insulating character of the system. Section VI contains a 
summary and discussion. 
 
\section{Strong-coupling limit} 
 
We begin by expressing the Hamiltonian for the AB$_{N-1}$ chain in a 
form to be used consistently throughout the following sections, 
\begin{eqnarray} 
H & = & -t \sum_{i\sigma} (c_{i+1\sigma}^{\dagger} c_{i\sigma} + {\rm H.c.}) 
 + U \sum_{i} n_{i\uparrow} n_{i\downarrow} \nonumber \\ & & \quad + \sum_{i}
\Delta_{i} n_{i}.  \label{ihm} 
\end{eqnarray} 
Here $c_{i\sigma}^{\dagger}$ creates an electron at site $i$ with spin  
$\sigma $, $n_{i\sigma} = c_{i\sigma}^{\dagger} c_{i\sigma}$, and  
$n_{i} = n_{i\uparrow} + n_{i\downarrow}$. The nearest-neighbor hopping 
amplitude is denoted by $t$, $U$ is the on-site Hubbard interaction and 
$\Delta_{i} = - \Delta$ if $i$ is a multiple of $N$ and zero otherwise. 
Thus $\Delta > 0$ is the difference in on-site energies between 
``metallic'' (B or $M$) and ``halogen'' (A or $X$) sites on an AB$_{N-1}$ 
($M_{N-1}X$) chain. 
 
A significant part of the physics of the AB$_{N-1}$ chain can be 
understood by considering the limit of small $t$. However, as in the 
case of the conventional IHM\ ($N = 2$), this limit presents particular 
complications for $U = \Delta $. Although it is possible at this special 
point to make a canonical transformation retaining three states per 
site,\cite{poldi} the resulting Hamiltonian is not trivial and no 
definitive conclusions may be drawn from it without further numerical 
analysis. This is beyond the scope of the current investigation, and 
in the remainder of this section we assume that $|U - \Delta | \gg t$. 
 
\begin{figure}[t!] 
\center \includegraphics[width=6.5cm,angle=0]{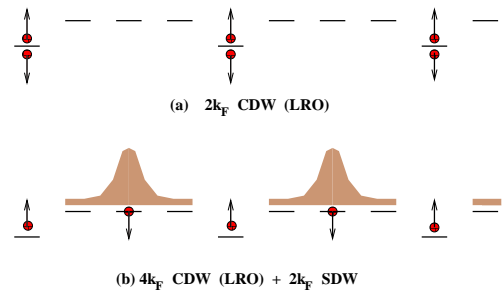} 
\caption{Charge and spin distribution in the strong-coupling limit, 
illustrated for an AB$_3$ chain.} 
\label{chd} 
\end{figure}

For $U < \Delta$, in a treatment discarding terms of order $[t/(\Delta 
- U)]^2$ the A sites are doubly occupied and the B sites are empty. 
Thus the charges are ordered in a charge-density wave (CDW) as represented 
in Fig.~\ref{chd}(a). The physics of this state can be described in terms 
of a one-particle picture in which the lower band is filled and the others 
are empty, yielding a BI. Again neglecting corrections of order $t^{2}$, 
the charge and spin gap are both equal to $\Delta - U$. We note that in 
contrast to the case $N = 2$, for $N > 2$ the charge distribution is 
altered radically if $\Delta$ is negative, because if $t \ll U$ the 
double occupancy at any site is no longer of order 1, but of order 
$(t/U)^2$, but here we will not consider this situation further. 
 
Another simple limit is $\Delta = 0$, where from the physics of the Hubbard 
model it is known that for any value of $t$ the system is a MI for $N = 2$ 
but a Luttinger liquid for any $N > 2$.\cite{lieb,schulz,ess} 

A further known result concerns the case $N = 2$ and $U - \Delta \gg t$, 
where the system adopts a form of modified MI state, which we will call 
here the ``correlated insulator'' (CI), with almost exactly one particle 
per site and a charge gap of approximately $U - \Delta$. This state has 
gapless spin excitations and long-ranged CDW order.\cite{brune,ali}
In spite of the fact that sites A and B are not equivalent, 
the system can be described by an effective Heisenberg model which is 
invariant under a one-site translation interchanging sites A and 
B.\cite{nag,ali} However, the expectation values calculated with 
this effective Hamiltonian are not invariant under the same one-site 
translation due to the accompanying mapping of the operators.\cite{ali} 
The charge difference between A and B sites (the amplitude of the CDW)
is at lowest order $22.2t^{2}\Delta U/(U^{2}-\Delta ^{2})^{2}$ and the 
decay of the charge-charge correlation function with distance $d$ has 
the form $d^{-3}\ln ^{-3/2}d$ for large $d$, with a prefactor proportional 
to $t^{4}\Delta ^{2}$.\cite{ali,man}
 
In order to extract analogous results for the case $N > 2$, we begin by 
considering the limit $U \rightarrow + \infty$, where the Hamiltonian for 
any $N$ can be mapped to a non-interacting spinless system with an effective 
magnetic flux.\cite{caspers,schad} For $N > 2$ and any positive $\Delta$, 
the system is an insulator with a charge distribution of the form depicted 
in Fig.~\ref{chd}(b). We will refer to this state as a CI to distinguish 
it from the charge distribution of the unperturbed MI. The primary 
differences from the case of the CI for $N = 2$ are that a finite value 
of $\Delta$ is required to drive the system into an insulating state, and 
that the gap in this state is very small. For $U \rightarrow + \infty$ and 
$N > 2$, this gap is given by  
\begin{eqnarray} 
E_{c} & = & 2 t \{\cos [\pi/N] - \cos [2\pi/N]\} \text{ if } \Delta \gg t,  
\nonumber \\ 
E_{c} & = & 2 \Delta/N \text{ \qquad if } \Delta \ll t \text{.}  \label{gapsc} 
\end{eqnarray} 
For $N > 2$ and $U - \Delta \gg t$ but with finite $U$, the spin degrees of 
freedom are important. A canonical transformation which eliminates doubly 
occupied sites maps the model into a generalized $t$-$J$ model, 
\begin{eqnarray} 
H & = & - t \sum_{i\sigma} P (c_{i+1\sigma}^{\dagger} c_{i\sigma} + 
{\rm H.c.}) P + \sum_{i} \Delta_{i} n_{i} \nonumber \\ 
 & & \quad + \sum_{i\delta = \pm 1} t_{i}^{ch} P [{c}_{i+\delta 
\sigma}^{\dagger} {c}_{i-\delta \sigma} (2 {\bf s}_{i} \cdot {\bf 
s}_{i-\delta } - {\textstyle \frac{1}{2}} n_{i})] P  \nonumber \\ 
 & & \quad + \sum_{i} J_{i} ({\bf s}_{i} \cdot {\bf s}_{i+1} - {\textstyle 
\frac{1}{4}} n_{i} n_{i+1}), 
\label{hsc} 
\end{eqnarray} 
where $P = \prod_{i} (1 - n_{i\uparrow} n_{i\downarrow})$ is the projector 
on the subspace of no double occupancy. The exchange parameter is given by 
\begin{eqnarray} 
J_{i} & = & \frac{4t^{2}U}{U^{2} - \Delta^{2}} \text{ for } i = lN \text{ 
or } i = lN-1, \nonumber \\ 
\text{ } J_{i} & = & \frac{4t^{2}}{U} \text{ otherwise,}  \label{j} 
\end{eqnarray} 
where $l$ is an integer, and the correlated hopping term by 
\begin{eqnarray} 
t_{i}^{ch} & = & \frac{t^{2}}{U - \Delta} \text{ for } i = lN \text{, (} 
l \text{ integer),}  \nonumber \\ 
t_{i}^{ch} & = & \frac{1}{2} \left( \frac{t^{2}}{U} + \frac{t^{2}}{U+\Delta}
\right) \text{ for } i = lN+1 \text{ or } i = lN-1,  \nonumber \\ 
t_{i}^{ch} & = & \frac{t^{2}}{U} \text{ otherwise.}  \label{ch} 
\end{eqnarray} 
 
As in the Hubbard model,\cite{oga} for the effective model of Eq.~(\ref{hsc}) 
the charge degrees of freedom are independent of the spin of the particles 
in the limit $U \rightarrow +\infty$, and can be described by a single 
Slater determinant corresponding to spinless fermions. The terms 
proportional to $J_{i}$ and $t_{i}^{ch}$ can be treated as a perturbation, 
and the resulting effective Hamiltonian $H_{s}$ for the spin degrees of 
freedom takes the form of a Heisenberg model for a ``squeezed'' chain, 
by which is meant one in which the empty sites are eliminated and the 
number of sites is equal to the number of electrons. The effective model 
$H_{s}$ is similar to generalized $t$-$J$ models which have been studied 
previously,\cite {pru,penc} with one important exception: because $J_{i}$ 
and $t_{i}^{ch}$ depend on the site, and the electrons which carry the spin 
are mobile (filling less than 1/2 for $N > 2$), the effective exchange term 
depends on the charge configuration. However, because the effective hopping 
$t$ is much larger than $J_{i}$ and $t_{i}^{ch}$ for $U - \Delta \gg t$ 
({\it i.e.}~the charge velocity is much larger than the spin velocity), 
it is reasonable to assume that all spin degrees of freedom experience 
an average effective exchange interaction. 
 
Thus one may write 
\begin{equation} 
H_{s} = \sum_{i} J_{\rm eff} ({\bf s}_{i} \cdot {\bf s}_{i+1} - {\textstyle 
\frac{1}{4}}), 
\label{hs} 
\end{equation} 
where 
\begin{eqnarray} 
J_{\rm eff} & = & \frac{1}{N} \sum_{i=0}^{N-1} [J_{i} \langle n_{i} n_{i+1} 
\rangle^{\prime} + 2 t_{i}^{ch} \langle {c}_{i-1}^{\dagger} {c}_{i+1} 
n_{i} \rangle^{\prime } \nonumber \\ 
 & & \quad + \, 2 t_{i+1}^{ch} \langle {c}_{i+2}^{\dagger}{c}_{i} n_{i+1} 
\rangle^{\prime}],  \label{jeff} 
\end{eqnarray} 
and the expectation values $\langle O \rangle^{\prime}$ of the operator  
$O$ are evaluated in the spinless model. It is well known that $H_{s}$ 
has a gapless spectrum with algebraic decay of the spin-spin correlation 
function and only short-ranged antiferromagnetic order, as represented 
in Fig.~\ref{chd}(b). 

In the effective Hamiltonian of Eq.~(\ref{hs}) we have included only 
terms up to order $t^2$ which break the spin degeneracy of the 
$U \rightarrow + \infty$ limit. However, a next-nearest-neighbor exchange 
interaction $J'$ appears at fourth order in $t$ in the conventional IHM 
($N = 2$),\cite{nag} and as $U$ is decreased the quantity $J'$ increases 
faster than does $J_{\rm eff}$. 
It is known that a spin gap opens for $J'/J > 0.2411 \dots$ in this spin 
model,\cite{nom} and thus a transition to a spin-gapped phase is expected 
as $U$ is lowered.\cite{brune} One then expects that for $N > 2$ the 
presence of terms of higher order in $t$ which are not included in 
Eq.~(\ref{hs}) also drive the same sort of spin transition as $U$ is 
lowered away from the strong-coupling limit.

To summarize the results of this section, from a strong-coupling analysis 
of the AB$_{N-1}$ chain one may conclude that, for fixed, finite $\Delta$ 
and small $U$, in the limit $t \rightarrow 0$ the system has a BI phase. 
As $U$ increases, the spin gap decreases within the BI regime and vanishes 
near $U \sim \Delta$, as in the conventional IHM. For higher values of $U$ 
the system adopts a CI phase which is similar to the MI in that charge 
degrees of freedom are high-lying and an effective spin model (albeit 
residing on a CDW background) describes the low-energy physics. However, 
from these considerations it is not possible to deduce the presence or 
absence of an intermediate FI phase. 
 
\section{Weak-coupling limit} 
 
In this section we analyze the model of Eq.~(\ref{ihm}) using bosonization, 
a technique which is applicable in the limit $(U,\Delta )\ll t$. A 
conventional weak-coupling analysis implies a small interaction $U$, but 
in this case one may not expand around the non-interacting case ($U = 0$) 
with $\Delta > 0$ because in this regime the system is a BI\ and has no 
Fermi points. Instead, for $U = \Delta = 0$ the AB$_{N-1}$ model reduces 
to a non-interacting chain with band filling $1/N$ (two particles in $N$ 
sites), for which the Fermi wave vector and Fermi velocity are given by  
\begin{equation} 
k_{F} = \frac{\pi}{N a_{0}} \text{ and } v_{F} = 2 t a_{0} \sin (a_{0} 
k_{F}),  \label{kf} 
\end{equation} 
respectively, where $a_{0}$ is the lattice spacing. 
 
As in the conventional IHM ($N = 2$), the bosonized expression for the 
$\Delta$ term [Eq.~(\ref{bhcs})] is strongly relevant, and renders the 
usual renormalization-group treatment of the model \cite{gog} invalid. 
For $N = 2$ some approximate and phenomenological treatments have 
predicted the existence of the FI phase and its fractionally charged 
excitations.\cite{fab,fab2,poldi} In particular, by starting from the 
CI phase and integrating approximately over the charge degrees of freedom, 
Fabrizio {\it et al}.~predicted that the spin transition takes place at a 
larger value of the on-site repulsion ($U_{s}$) than does the charge 
transition ($U_{c}$).\cite{fab} Taking the effective potential obtained 
from bosonization as a phenomenological Ginzburg-Landau free energy, these 
authors found that the excitations in the FI phase have a fractional 
charge which varies between 1 at the boundary with the BI phase and 0 at 
the boundary with the CI phase. Thus the elementary excitation interpolates 
between an electron in the BI phase and a spinon in the CI phase, and its 
fractional charge is proportional to the electric polarization in the FI 
phase.\cite{poldi} These excitations may be visualized as the topological 
excitations of an effective spin-1 chain for $t \ll U, 
\Delta$.\cite{dimer,poldi} 
 
In the same spirit as the analysis of Fabrizio {\it et al.}, we identify 
the most important operators for $N > 2$ chains on the basis of their 
critical dimensions and their expected effect on the physics. From these 
we infer the qualitative features of the phase diagram, which will be 
compared with numerical results in Sec.~V. 
 
We linearize the spectrum and pass to the continuum limit by the substitution  
\begin{equation}
c_{j\sigma} \rightarrow \sqrt{a_{0}} \left[ \mbox{e}^{{\it i} k_{F} x}
\Psi_{+,\sigma}(x) + \mbox{e}^{-{\it i} k_{F} x} \Psi_{-,\sigma }(x) \right]\,,
\label{lineari}
\end{equation}
where the original lattice operators are decomposed into right- and 
left-moving components $\Psi_{+,\sigma}^{\dagger}(x)$ and $\Psi_{-,
\sigma}^{\dagger}(x)$, and $x = ja_{0}$. Bosonization of these fields 
by a standard method for electrons with spin\cite{voit,Senechal} yields
\begin{equation}
\Psi_{\pm,\sigma}^{\dagger}(x) \simeq \frac{\kappa_{\pm,\sigma}^{\dagger}}
{\sqrt{2 \pi a_{0}}} \exp (\mp i\phi_{\pm \sigma} + \theta_{\pm \sigma }),  
\label{rlbos}
\end{equation}
where $\phi_{+\sigma}$ ($\phi_{-\sigma}$) are right- (left)-moving Bose 
fields and $\theta_{r\sigma}$ is the field dual to $\phi_{r\sigma}$. The 
Klein factors $\kappa_{r\sigma}^{\dag}$ have the physical meaning of ladder 
operators which increase by one the number of fermions in branch $r$ with 
spin $\sigma$, and also serve to ensure that the anticommutation relations 
for electron fields of different spin are maintained. They are Hermitian 
and satisfy a Clifford algebra\cite{note}
\bea
& \{\kappa_{+\sigma}, \kappa_{+\sigma^{\prime}} \} = \{\kappa_{-\sigma}, 
\kappa_{-\sigma^{\prime}}\} = 2 \delta_{\sigma \sigma^{\prime}}\, & 
\nonumber \\ & \{\kappa_{+\sigma}, \kappa_{-\sigma^{\prime}} \} = 0. &  
\label{klein}
\eea
We define
\[
\phi_{\sigma} = \phi_{-\sigma} + \phi_{+\sigma}, \qquad \theta_{\sigma} 
= \phi_{-\sigma } - \phi _{+\sigma },\
\]
and introduce the linear combinations 
\begin{equation}
\phi_{c} = {\textstyle \frac{1}{\sqrt{2}}} (\phi_{\uparrow} 
+ \phi_{\downarrow}), \qquad
\theta_{c} = {\textstyle \frac{1}{\sqrt{2}}} (\theta_{\uparrow} 
- \theta_{\downarrow })
\label{bos_carge}
\end{equation}
\begin{equation}
\phi_{s} = {\textstyle \frac{1}{\sqrt{2}}} (\phi_{\uparrow} 
+ \phi_{\downarrow}),\qquad
\theta_{s} = {\textstyle \frac{1}{\sqrt{2}}} (\theta_{\uparrow} 
- \theta_{\downarrow})
\label{bos_spin}
\end{equation}
to describe respectively the charge ($\phi_{c}$) and spin ($\phi_{s}$) 
degrees of freedom and their conjugate momenta.

With the exception of the term in ${\cal A}_{4k_{F}}$ below, the above 
equations combined with appropriate operator product expansions give the 
bosonized expressions for the charge density  
\begin{eqnarray}
\rho_{c}(x) & \equiv &: n_{i\uparrow} + n_{i\downarrow}: \;\; \rightarrow \; 
{\textstyle \frac{1}{\sqrt{2}\pi}} (\partial_{x} \phi_{c}) \label{cd} \nonumber \\
& & + {\cal A}_{2k_{F}} \cos (2 k_{F} x + \sqrt{2} \phi_{c}) \cos 
\sqrt{2} \phi_{s} \nonumber \\ 
& & - {\cal A}_{4k_{F}} \cos (4 k_{F} x + \sqrt{8} \phi_{c}), 
\end{eqnarray}
and for the spin density
\begin{eqnarray} 
\rho_{s}(x) & \equiv & :{\textstyle \frac{1}{2}} ( n_{i\uparrow} - 
n_{i\downarrow}): \;\; \rightarrow \; {\textstyle \frac{1}{\sqrt{2} \pi}} 
(\partial_{x} \phi_{s}) \label{sd} \nonumber \\
& & \, + {\cal B}_{2k_{F}} \cos (2 k_{F} x + \sqrt{2} \phi_{c}) \sin \sqrt{2} 
\phi_{s} .  
\end{eqnarray}
The term in ${\cal A}_{4k_{F}}$, which is crucial in our analysis, is not 
present in a standard bosonization procedure but is generated in any model 
with density-density interaction terms. This may be seen by considering the lowest-order 
correction in the dressed charge operator, which includes terms of the 
form $e ^{- 4 i k_F} \Psi_{+,\sigma}^{\dagger} \Psi_{+,\sigma}^{\dagger} 
\Psi_{-,\sigma} \Psi_{-,\sigma}$, where each electron operator 
$c_{j\sigma}^{\dag}$ contributing to the four-fermion product 
introduces a factor of $e^{- i k_F}$ [see Eq.~(\ref{lineari})]. 
The presence of this term has been established definitively in the 
Hubbard model (any AB$_{N-1}$ chain with $\Delta = 0$).\cite{bed} 

The coefficients ${\cal A}_{2k_{F}}$, ${\cal A}_{4k_{F}}$, and ${\cal 
B}_{2k_{F}}$ are non-universal parameters which depend on $U$ and fulfil 
the conditions  
\begin{eqnarray} 
\lim_{U \rightarrow 0} \,\, {\cal A}_{2k_{F}}(U) & = & {\cal A}_{1}^{0} = 
{\textstyle \frac{1}{2\pi}} \, ,  \nonumber \\ 
\lim_{U \rightarrow 0} \,\, {\cal B}_{2k_{F}}(U) & = & {\textstyle 
\frac{1}{2}} {\cal A}_{1}^{0} \, ,  \nonumber \\ 
\lim_{U \rightarrow 0} \,\, {\cal A}_{4k_{F}}(U) & = & 0\, ,  \nonumber \\ 
\lim_{U \rightarrow + \infty} \,\, {\cal A}_{4k_{F}} (U) & = & {\cal 
A}_{1}^{0} \, .
\label{limitb1} 
\end{eqnarray} 
The coefficient ${\cal A}_{4k_{F}}$ has been calculated numerically for 
the particular case of the 0.55-filled Hubbard chain.\cite{bed} It has 
a monotonic behavior, linear in $U$ for $U \rightarrow 0$ and with a 
downward curvature. For $U = 10t$ it is of order 0.12, already close to 
its saturation value ${\cal A}_{1}^{0} \simeq 0.159$ for $U \rightarrow 
+ \infty$. For the filling $1/N$ under consideration [two particles per 
unit cell, Eq.~(\ref{kf})], a $2k_{F}$ modulation corresponds to the 
periodicity of the AB$_{N-1}$ lattice, while $4k_{F}$ corresponds to a 
modulation of half this period in real space (Fig.~\ref{chd}). 
 
To derive the bosonized expression of the on-site energy term in the 
continuum limit, we note that it can be written in the form 
\begin{equation} 
\sum_{j} \Delta_{j} n_{j} = - \frac{\Delta}{N} \sum_{j} \sum_{l=0}^{N-1} 
\exp (2 i k_{F} a_{0} l j) n_{j}.  \label{onsite} 
\end{equation} 
By transforming the sum over $j$ into an integral over $x = j a_{0}$, using 
Eq.~(\ref{cd}), and neglecting rapidly oscillating factors (which here means 
retaining only those terms in $\exp (ik_{F} a_{0} n)$ with $n$ a multiple of 
$2N$), we obtain  
\begin{eqnarray} 
\sum_{j} \Delta_{j} n_{j} & \rightarrow & -\frac{\Delta}{N} 
{\cal A}_{2k_{F}} \int dx \, \cos (\sqrt{2} \phi _{c}) \cos (\sqrt{2} 
\phi_{s})  \nonumber \\ 
& & + \frac{\Delta}{N} \, {\cal A}_{4k_{F}} \int dx \, \cos (\sqrt{8} 
\phi_{c}). \label{dv1} 
\end{eqnarray} 
For the interaction term, with $N > 2$ one obtains the usual form 
\begin{eqnarray} 
U \sum_{i} n_{i\uparrow} n_{i\downarrow} & \rightarrow & \int dx \left\{ 
\frac{U}{2\pi^{2}} [(\partial_{x} \phi_{c})^{2} - (\partial_{x} 
\phi_{s})^{2}] \right. \nonumber \\ & & \quad + \, \left. \frac{U}{(2 \pi 
a_{0})^{2}} \cos (\sqrt{8} \phi_{s}) \right\}. 
\label{huve1p} 
\end{eqnarray} 
Only for $N = 2$ does one have in addition the contribution from Umklapp 
scattering processes
\begin{equation} 
{\cal H}_{\rm Um} = \frac{U}{(2\pi a_{0})^{2}} \int dx \cos (\sqrt{8} 
\phi_{c}),  
\label{umk} 
\end{equation} 
which has the same form as the last term in Eq.~(\ref{dv1}). We note that,
even at weak coupling where the bosonization results are applicable, 
$U \gg \Delta {\cal A}_{4k_{F}}$ and ${\cal H}_{\rm Um}$ is the dominant 
term in the $N = 2$ case.\cite{fab}
 
On including the non-interacting part, the final Hamiltonian density for 
the bosonized model can be expressed as 
\begin{eqnarray} 
{\cal H}_{{\rm eff}} & = & {\cal H}_{c} + {\cal H}_{s} + {\cal H}_{cs} \, , 
\label{bh} \\ 
{\cal H}_{c} & = & \frac{\pi v_{c} K_{c}}{2} \Pi_{c}^{2} (x) + \frac{v_{c}}
{2 \pi K_{c}}(\partial_{x} \phi_{c})^{2}  \nonumber \\ 
& & \quad + \frac{M_{c}}{(2\pi a_{0})^{2}} \cos \left( \sqrt{8} \phi_{c} 
(x) \right) , \label{bhc} \\ 
{\cal H}_{s} & = &\frac{\pi v_{s}}{2} \Pi_{s}^{2}(x) + \frac{v_{s}}{2 \pi}
(\partial_{x} \phi_{s})^{2}] \nonumber \\ 
& & \quad + \frac{M_{s}}{(2 \pi a_{0})^{2}} \cos \left( \sqrt{8} \phi_{s} 
(x) \right) ,  \label{bhs} \\ 
{\cal H}_{cs} & = & - \frac{m_{cs}}{\pi a_{0}} \cos \left( \sqrt{2} \phi_{c}
\right) \cos \left( \sqrt{2} \phi_{s} \right) .  \label{bhcs} 
\end{eqnarray} 
Here $\Pi_{\nu}$ is the moment conjugate to $\phi_{\nu}$, $v_{c}$ 
and $v_{s}$ are the charge and spin velocities, and $K_{c}$ is the 
charge-correlation exponent. For small interactions, the values of 
the different parameters are  
\begin{eqnarray} 
M_{c} & \simeq & U \qquad \qquad \qquad \;\; \text{if } N = 2,  \nonumber \\ 
M_{c} & \simeq & \Delta \cdot {\cal A}_{4k_{F}}(U) \qquad \text{if } N > 2,  
\nonumber \\ 
K_{c} & \simeq & 1 - U a_{0}/\pi v_{F}, \quad \; v_{c},v_{s} \simeq v_{F} 
\nonumber \\ 
M_{s} & \simeq & U, \qquad\qquad\qquad \; m_{cs} \simeq \Delta .  
\label{mcs} 
\end{eqnarray} 
This Hamiltonian density coincides formally with the field theory studied 
previously \cite{fab,fab2} for the case $N = 2$ corresponding to the IHM, 
with the sole difference arising in the amplitude of the effective Umklapp 
scattering term $M_{c}$. 

To discuss propeties of the different phases which appear in the AB$_{N-1}$ 
model, in addition to the on-site charge and spin density operators defined 
in Eqs.~(\ref{cd}) and (\ref{sd}) we will use the bosonized expressions 
for the on-bond charge and spin density operators,\cite{voit2,Japaridze_1995}
\begin{eqnarray}
B_{i} & = & \sum_{\sigma} (c_{i,\sigma}^{\dagger} c_{i+1,\sigma} + 
c_{i+1,\sigma}^{\dagger} c_{i,\sigma} )   \nonumber \\ & \rightarrow & 
{\textstyle \frac{1}{2 \pi}} [ \Pi_c^2 (x) + (\partial_x \phi_c)^2] 
\nonumber \\ & & + \cos ( (2i+1)\pi /N + \sqrt{2} \phi_{c} ) \cos \sqrt{2} 
\phi_{s},  \label{ob} \\
W_{i} & = & \sum_{\sigma} \sigma ( c_{i,\sigma}^{\dagger} c_{i+1,\sigma}
 + c_{i+1,\sigma}^{\dagger} c_{i,\sigma})  \nonumber \\ & \rightarrow & 
{\textstyle \frac{1}{2 \pi}} [ \Pi_s^2 (x) + (\partial_x \phi_s)^2] \nonumber 
\\ & & + \cos ( (2i+1)\pi /N + \sqrt{2} \phi_{c}) \sin \sqrt{2} \phi_{s}.  
\label{ow}
\end{eqnarray}
To characterize the FI phase we introduce the order parameter 
\[
\mathcal{O}_{\mathrm{FI}}=\sum_{j}\cos (2k_{F}a_{0}j)(B_{j}-B_{j-1}),
\]
which represents the $2k_{F}$ Fourier component of the 
difference between consecutive bond-density operators, 
and whose bosonized expression takes the form 
\begin{equation}
\mathcal{O}_{\mathrm{FI}} \simeq \sin (\pi/N) \sin (\sqrt{2} \varphi_{c}) 
\cos (\sqrt{2} \varphi_{s}).  \label{obow}
\end{equation}
We note that the operator $\mathcal{O}_{\mathrm{FI}}$ is antisymmetric 
with respect to inversion at the A sites ($j \rightarrow - j$).

Classically, it is clear that minimization of the ionic term in 
Eq.~(\ref{bhcs}) requires either $\phi_{c} = \phi_{s} = 0$ or $\sqrt{2}
\phi_{c} = \sqrt{2} \phi_{s} = \pi$ (mod $2\pi$). These values of $\phi_{c}$ 
and $\phi_{s}$ characterize the BI phase, as they ensure that bond-order
parameter $\mathcal{O}_{\mathrm{FI}}$ and the site and bond spin densities
[Eqs.~(\ref{sd}) and (\ref{ow})] are suppressed, while there remains a 
long-ranged CDW order at $2 k_F$ [Eq.~(\ref{cd})] with increased charge 
on the A atoms (for which $x$ is multiple of $N a_{0}$). The amplitude 
of the $4k_{F}$ CDW is much smaller because it is proportional to 
$\mathcal{A}_{4k_{F}}$. We stress that due to the symmetry of the 
Hamiltonian a $2k_{F}$ CDW is present in all phases of the model. In 
contrast to the $N = 2$ case there is also a modulation of the bond order, 
or bond-order wave, in the BI phase with $B_{i} \propto \cos [(2i+1) 
\pi /N]$ (\ref{ob}) as a consequence of the $N$-site unit cell. This
distribution of bond intensities is represented in Fig.~2(a), and is 
symmetric under reflection at the A sites ($\mathcal{O}_{\mathrm{FI}} 
= 0$).

\begin{figure}[t!] 
\center \includegraphics[width=6.5cm,angle=0]{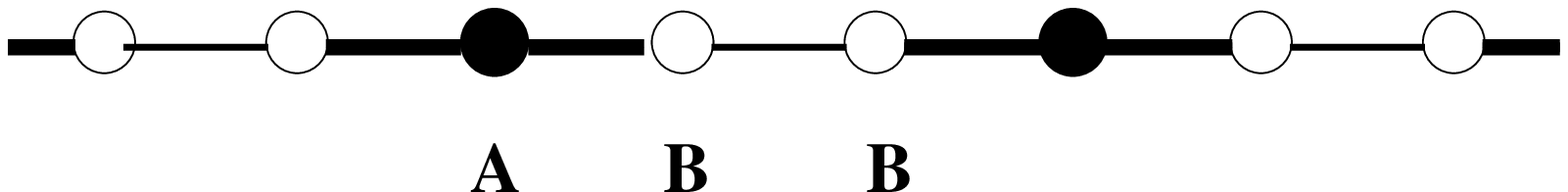} 
\centerline{(a)}
\center \includegraphics[width=6.5cm,angle=0]{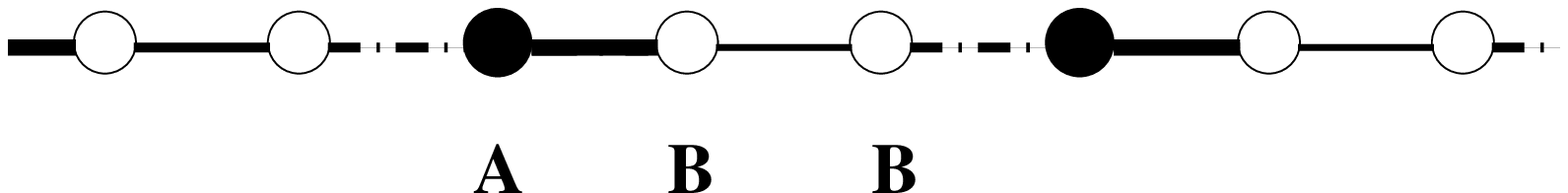} 
\centerline{(b)}
\caption{Bond-density distribution (a) in the BI phase and (b) in the FI 
phase of an AB$_2$ chain. Thick lines correspond to high on-bond density, 
thin lines to less occupied bonds, and dashed lines to bonds with lowest 
occupation.}
\label{Fig2} 
\end{figure} 

While $\sqrt{8} \phi_{c} = 0$ (mod $2\pi$) in the BI phase, minimizing the
potential of $\mathcal{H}_{c}$ [Eq.~(\ref{bhc})] requires $\sqrt{8} \phi_{c}
= \pi $ (mod $2\pi$). Thus for fixed $\Delta$ there is a charge transition
with increasing $U$ for $N = 2$.\cite{fab,fab2} For $N > 2$, $M_{c}$ is
determined not only by the on-site repulsion $U$ but also by the ionicity
parameter $\Delta $. At small $U$, $M_{c} \simeq \Delta U/t$, but for large 
$U$ one expects $M_{c} \simeq \Delta$ as a result of the saturation of 
$\mathcal{A}_{4k_{F}}(U)$ in the Hubbard model (above). In consequence it 
is not possible to assure, as in the case $N = 2$, that this effect will
overcome the ionicity for sufficiently large on-site repulsion, leading to a
transition in the charge sector. However, from the results of the previous
section, in the strong-coupling limit ($t \rightarrow 0$) there is a charge
transition for $U = U_{c}$ with $U_{c} \sim \Delta$. There is also a spin
transition at $U = U_{s}$ because the BI phase has a spin gap which vanishes
when $U - \Delta \gg t$. Thus it appears that the same generic phase diagram
indeed emerges for all values of $N$.

For $N = 2$, the authors of Ref.~\onlinecite{fab} argued that $U_{s} > U_{c}$ 
by describing the charge degrees of freedom in the CI phase in terms of a free
boson of mass (charge gap) $E_{c}$. On integrating over the charge degrees
of freedom, the resulting effective Hamiltonian takes the form of Eq.~(\ref
{bhs}) but with an effective renormalized mass 
\begin{equation}
M_{s}^{\prime } - M_{s} \simeq -8 \pi t(\Delta / E_{c})^{2},  \label{rm}
\end{equation}
whence $U_{s}$ is determined by the condition $M_{s}^{\prime} = 0$. For 
$U_{c} < U < U_{s}$ the system retains a spin gap as in the BI phase, with 
the values $\varphi_{s} = 0$ or $\varphi_{s} = \pi /\sqrt{2}$ frozen, but 
the nature of the charge sector has changed. The spin correlations still 
decay exponentially due to the presence of the spin gap.

An analysis of the intermediate regime may be performed for higher values 
of $N$ by analogy with that applied to the model with $N = 2$. This type of 
treatment\cite{fab,poldi} shows that the intermediate phase is characterized 
in the bosonized description by values of $\sqrt{8} \phi_{c}$ intermediate 
between 0 and $\pi$, and thus possesses fractional-charge excitations, broken
inversion symmetry, and the non-zero polarizability of a ferroelectric phase. 
Because $\phi_{c} \neq 0$, one observes from Eq.~(\ref{obow}) that the order 
parameter is finite, $\mathcal{O}_{\rm FI} \ne 0$, meaning that the charge 
distribution in the intermediate phase is characterized by broken inversion 
symmetry at the A sites. The long-ranged order of the BI phase, namely the 
$2 k_F$ modulation of the site charge density and the $N$-site-periodic 
modulation of the bond density, is retained. The distribution of bond 
charge density resulting from Eq.~(\ref{ob}) for one of the two possible 
inequivalent choices of $\phi_{\nu}$ minimizing the energy is represented 
in Fig.~2(b), which makes clear the spontaneous breaking of parity in the 
FI phase. 

Qualitatively, the properties of the FI phase may be understood in this
framework by taking the interaction part of the Hamiltonian density as a
phenomenological Ginzburg-Landau energy functional with effective
interactions.\cite{fab} In the notation chosen here,\cite{note} this energy
functional takes the form\cite{poldi} 
\begin{eqnarray}
F & = & M_{c} \alpha_{c}^{2} + M_{s} \alpha_{s}^{2} + m_{cs} \alpha_{c}
\alpha _{s},  \nonumber \\
\text{with } \alpha_{\nu} & = & \cos (\sqrt{2} \phi_{\nu}),  \label{f}
\end{eqnarray}
with $M_{s} < 0$, $M_{c} > 2m_{cs}$, and all parameters duly renormalized.
The minima of $F$ occur when $\alpha_{c} = m_{cs}/(2M_{c})$, $\alpha_{s} = -
1$, or $\alpha_{c} = - m_{cs}/(2M_{c})$, $\alpha_{s} = 1$. A soliton between
two segments of the system characterized by these pairs of values
corresponds to an elementary excitation of spin 1/2 and charge $C = 1 - 2
\arccos [m_{cs}/(2M_{c})]/\pi $ (\textit{i.e.}~proportional to the
difference between the two values of $\phi_{c}$). The polarization of the
homogeneous system turns out to be $\pm eC/2$.\cite{poldi}

Finally, for $U > U_{s}$, the spin gap is absent and the system is in the 
CI phase. With increasing on-site repulsion the amplitude of the $2k_{F}$ 
modulations of the on-site charge density, ${\cal A}_{2k_{F}}$, decreases 
and the $4k_{F}$ component becomes dominant. Because of the spatial 
modulation of $\Delta$ there is long-ranged CDW order in the ground 
state\cite{brune,ali} as when $N = 2$. However, on top of this order the 
system also possesses the antiferromagnetic spin ordering represented in 
Fig.~\ref{chd}(b). As a consequence of the gapless character of the spin 
excitations, all correlation functions decay with a power-law form, 
including fluctuations of the charge density because of the strong 
charge-spin coupling [Eq.~(\ref{bhcs})]. These correlation functions have 
been considered by one of us for the case $N = 2$.\cite{ali} 
 
We conclude this section by extracting from the form of the bosonized 
effective Hamiltonian (\ref{bh}) the following differences between the 
well-characterized AB system and the AB$_{N-1}$ chain with $N > 2$. 

\noindent
1) Because the amplitude of $M_{c}$ is very much smaller than for the 
AB chain, the BI gap decreases more slowly with increasing $U$ and the 
BI phase should extend to larger values of $U$.
 
\noindent
2) The charge gap $E_{c}$ in the CI phase ($U > U_{s}$) is determined not 
only by $U$ but also by the ionicity parameter $\Delta $. In particular, 
because the amplitude ${\cal A}_{4k_{F}}(U)$ of the $4k_{F}$ modulations 
saturates for large $U,$ one expects that $E_{c}$ is determined solely by 
$\Delta $ and $t$ for $U\rightarrow +\infty $, in agreement with the 
results of the previous section. 
 
\noindent
3) The smaller value of $E_{c}$ implies a larger renormalization of the 
effective spin mass $M_{s}^{\prime }$ [Eq.~(\ref{rm})], and thus the 
FI phase is expected over a wider parameter range. 
 
\section{Numerical methods} 

We have performed numerical calculations of a number of quantities 
characterizing the physical properties of AB$_{N-1}$ chains for 
comparison with the analytical considerations of Secs.~II and III, 
and to interpolate between the limits these represent. Here we 
describe the numerical techniques applied to the Hamiltonian of 
Eq.~(\ref{ihm}). 

We have determined numerically the phase diagram for systems of $L$ sites 
by considering the crossing of appropriate excited energy levels which 
correspond to jumps in the charge ($\gamma_{c}$) and spin ($\gamma_{s}$) 
Berry phases, and by computing the charge ($z_{L}^{c}$) and spin 
($z_{L}^{s}$) localization parameters defined below. We have in addition  
calculated the Drude weight $D_{c}$ in some cases as a supplementary 
probe of metallic behavior. 

The parameter $\gamma_{c}$ is the Berry phase captured by the ground state 
for a ring threaded by a flux which is varied adiabatically from zero to 
two flux quanta.\cite{res3,ort2,pola} The spin Berry phase $\gamma _{s}$ 
is the corresponding quantity for a situation in which oppositely directed 
fluxes are experienced by spin-up and spin-down particles.\cite{epl,bos,topo} 
Specifically, under an applied flux (which we have scaled to be dimensionless) 
$\Phi_{\sigma} h c/(2 \pi e)$ for spin $\sigma$, the Hamiltonian $H$ 
[Eq.~(\ref{ihm})] is transformed into a Hamiltonian $\widetilde{H}$ which 
differs from $H$ in that the hopping term has the form $- t \sum_{i\sigma} 
(\widetilde{c}_{i+1\sigma}^{\dagger} \widetilde{c}_{i\sigma} e^{i 
\Phi_{\sigma}/L} + {\rm H.c.}).$ We denote by $|g(\Phi_{\uparrow}, 
\Phi_{\downarrow })\rangle$ the ground state of $\widetilde{H} 
(\Phi_{\uparrow }, \Phi_{\downarrow })$. The charge (spin) Berry phase 
$\gamma_{c\text{ }}$($\gamma _{s}$) is the overall phase captured by 
the state $|g(\Phi_{\uparrow}, \Phi_{\downarrow}) \rangle$ on following 
adiabatically the cycle $0 \leq \Phi \leq 2\pi$ with $\Phi_{\uparrow} = 
\Phi_{\downarrow} = \Phi$ ($\Phi_{\uparrow} = - \Phi _{\downarrow } = 
\Phi$). Discretizing the interval $0 \leq \Phi \leq 2\pi $ into $N+1$ 
points $\Phi_{r} = 2 \pi r/N$ ($r = 0,N$), the Berry phases are calculated 
from the numerically gauge-invariant expression\cite{pola}  
\begin{eqnarray} 
\gamma_{c(s)} & = & \lim_{N \rightarrow \infty} \text{Im} \{ \ln 
[\prod_{r=0}^{N-2} \langle g(\Phi_{r},\pm \Phi_{r}) |g(\Phi_{r+1},\pm 
\Phi_{r+1}) \rangle \nonumber \\ 
& & \times \langle g(\Phi_{N-1},\pm \Phi_{N-1}) |g(2\pi ,\pm 2\pi ) 
\rangle ]\}.  \label{gamma} 
\end{eqnarray} 
The state $|g(2\pi, 2\pi) \rangle$ ($|g(2\pi,-2\pi) \rangle$) is obtained 
from $|g(0,0) \rangle$ by applying the displacement operator $U_{L}^{c}$ 
($U_{L}^{s}$), 
\begin{equation} 
|g(2\pi,\pm 2\pi) \rangle = U_{L}^{c(s)} |g(0,0) \rangle ,  \label{be2} 
\end{equation} 
with 
\begin{eqnarray} 
U_{L}^{c} & = & \exp \left[i \frac{2 \pi}{L} X \right] \text{, } X = 
\sum_{j} j a_{0} (n_{j\uparrow} + n_{j\downarrow}),  \label{ucl} \\ 
U_{L}^{s} & = & \exp \left[i \frac{2 \pi}{L} D \right] \text{, } D = 
\sum_{j} j a_{0} (n_{j \uparrow } - n_{j\downarrow }).  \label{usl} 
\end{eqnarray} 
Note that Eq.~(\ref{be2}) is simply a gauge transformation which transforms 
the ground state of $H$ into that of $\widetilde{H} (2\pi,\pm 2\pi )$. The 
operator $U_{L}^{c}$, the exponential of the total-position operator, 
constitutes a translation in momentum space which displaces all 
single-particle wave vectors by $2\pi /L$.\cite{znos} Similarly, 
$U_{L}^{s}$ displaces the wave vectors of up and down particles in 
opposite directions. 
 
An important property of the charge Berry phase is that if the system is 
modified by some perturbation, the change in polarization $P_{\uparrow}
 + P_{\downarrow}$ is proportional to the corresponding change in $\gamma 
_{c}.$\cite{ort2} Here $P_{\sigma}$ is the contribution of electrons with 
spin $\sigma $ to the polarization of the system. Similarly, changes in  
$\gamma_{s}$ are related to changes in the difference $P_{\uparrow} - 
P_{\downarrow}$ between the electric polarizations for up and down 
spins,\cite{epl}
\begin{equation} 
\Delta P_{\uparrow} \pm \Delta P_{\downarrow} = e \Delta \gamma_{c(s)}/2\pi  
\text{ } (\text{mod }e).  \label{po} 
\end{equation} 
The AB$_{N-1}$ chain has site inversion symmetry at the A sites for all 
$N$, and also at the B sites for $N = 2$. A crucial property for the 
purposes of the present analysis is that, in systems with site inversion 
symmetry, $\gamma_{c}$ and $\gamma_{s}$ can only take the values $0$ or 
$\pi$ (mod $2 \pi$) (the argument of the logarithm in Eq.~(\ref{gamma}) 
becomes its own complex conjugate under inversion). Thus the Berry-phase 
vector ${\bgamma} = (\gamma_{c},\gamma_{s})$ cannot vary 
continuously, and a jump in ${\bgamma}$ corresponds to a transition 
in at least one of the topological quantum numbers $\gamma_{c}/\pi$ and
$\gamma _{s}/\pi $. These topological transitions usually correspond to 
phase transitions in the thermodynamic limit. As an example, in the 
strong-coupling limit the CI phase of the IHM has one charge at every 
site (1111\dots), while the BI is characterized by a charge distribution 
of alternating electron pairs (2020\dots). In order to transform from one 
to the other it is necessary to displace half of the charges by one lattice 
parameter, which corresponds to a change $\pm e/2$ in the polarization of 
the system, and according to Eq.~(\ref{po}) the transition is accompanied 
by a jump of $\pi$ in $\gamma_{c}$. Similarly, it can be shown that the 
opening of a spin gap in a Luttinger-liquid phase of a spin-rotationally 
invariant model is accompanied by a topological transition in 
$\gamma_{s}$.\cite{epl} Extrapolating the parameters for which these 
jumps occur in systems with $L \leq 16$ to the thermodynamic limit has 
led to a very accurate determination of the quantum phase diagram of the 
Hubbard model with correlated hopping.\cite{bos,topo} Other successful 
applications of the method include the extended Hubbard chain with 
charge-dipole interactions\cite{pola} and the conventional IHM\cite{tor} 
($N = 2$ of the present case). 
 
In the limit in which all charges are localized (generally the 
strong-coupling limit), the Berry phases are easy to calculate 
analytically. For $t = 0$ one may choose a gauge in which all 
scalar products in Eq.~(\ref{gamma}) are equal to 1, except possibly 
the last. The Berry phases are then defined by the argument of 
the exponential of $U_{L}^{c(s)}$ in Eq.~(\ref{be2}). An illustrative 
example is the charge distribution of the BI for $t \rightarrow 0$ 
[Fig.~\ref{chd}(a)] in a system of $M$ unit cells each of length $N$, 
such that $L = MN$: it is clear that $D = 0$ (\ref{usl}) and thus 
$\gamma_{s} = 0$, while [see Eq. (\ref{ucl})]  
\[ 
X = \sum_{l=0}^{M-1} 2 N l = N M (M-1), 
\] 
whence $U_{L}^{c} = \exp [2 \pi i (M-1)] = 1$ and $\gamma_{c} = 0$ 
(mod $2 \pi$). By contrast, for a system of particles with spin up 
localized at sites $0,N,2N,\dots$ and those with spin down localized 
at positions $N/2,N+N/2,2N+N/2,\dots$ [Fig.~\ref{chd}(b)], $X = NM(M-1)
+ NM/2$ and $D = -NM/2$, leading to $U_{L}^{c} = U_{L}^{s} = -1$ and 
$\gamma_{c} = \gamma_{s} = \pi$; these values characterize the CI. 

The jumps in the Berry phases coincide in general with the crossing
of pairs of low-lying excited states of a finite system. Thus the 
critical values $U_c$ and $U_s$ may be determined by the method of
crossing excitation levels (MCEL), where this method is construed in 
a broader sense to be defined below. The relevant level crossing is 
found where the energy of the ground state $|g(\Phi, \pm \Phi) \rangle$
as a function of the flux $\Phi$ reaches its maximum value.\cite{tor}
This is usually equivalent to the situation obtained by taking boundary
conditions (BCs) opposite to the ``closed-shell'' choice which leads to the
minimum energy.\cite{tor} Identifying the two crossing levels by their
quantum numbers and finding the parameters for which their energies
coincide is naturally equivalent to locating the jump in the Berry
phase, and the former procedure saves computer time by avoiding the
evaluation of Eq.~(\ref{gamma}). Thus we have followed this method in 
determining the majority of the data to be shown in Sec.~V, although 
we have also used Eq.~(\ref{gamma}) for verification of our results.

In its more restricted sense, the MCEL is based on identifying the
appropriate levels by conformal field theory with renormalization-group
analysis.\cite{nom,nak,somma} This is a weak-coupling approach which takes
advantage of the fact that in conformally invariant systems of finite size
the smallest excitation gap corresponds to the dominant correlations at
large distances. In previous studies, which included the Hubbard model 
with correlated hopping,\cite{bos,topo,nak} the extended Hubbard
model,\cite{tor,nak} and the conventional IHM,\cite{tor} the restricted 
MCEL and the jumps of Berry phases were found to coincide, giving support 
to the method from both weak- and strong-coupling analyses.

From the arguments presented in the strong-coupling limit, it is evident
that the jump in $\gamma_{c}$ characterizes a sharp transition with a
reordering of charge. One might expect by continuity that this jump
characterizes the charge transition also at weak coupling. While in 
previous studies (above) the jump in $\gamma_{c}$ does coincide with the 
crossing of appropriate excitations determined from field-theoretical 
arguments,\cite{tor,nak} for larger $N$ it is not known whether the level 
crossing which corresponds to the jump in $\gamma_{c}$ has support from a 
weak-coupling approach. In the following, the MCEL refers more broadly 
to a crossing which corresponds to the jump in a Berry phase, but in the case 
of $\gamma_{c}$ this correspondence is not justified by field-theoretical 
methods. When considering $\gamma_{s}$ in any model with SU(2) symmetry, this 
crossing always coincides with that corresponding to the Kosterlitz-Thouless 
transition for the opening of a spin gap according to conformal field 
theory.\cite{epl}

The specific crossing levels in the MCEL are determined by the quantum 
numbers specifying their parity, spin, and total momentum, plus the BCs 
for which the crossing occurs. The spin transition is determined from the 
crossing of an even singlet and an odd triplet excited state, both with 
total momentum $K = 0$, when working with open-shell BCs. This is the 
crossing which corresponds to the jump in the spin Berry phase\cite{epl} 
and has support from field-theoretic arguments.\cite{nom,nak} The charge 
transition is determined from the crossing of an even and an odd singlet, 
both with $K = \pi/a_0$, again calculated with open-shell BCs. This crossing 
signals the jump in the charge Berry phase, which as explained above denotes 
a reordering of charge leading to a jump in the polarization of the system. 
However, as in other recent calculations in which crossing of levels was 
used to determine phase transitions in two-dimensional spin 
systems,\cite{sind,nic} a systematic demonstration that this procedure 
should remain valid in the weak-coupling limit is lacking. For this reason, 
and because of other shortcomings described below, we reinforce the results 
obtained from the MCEL by comparison with those obtained from the localization 
operators.

The localization operators are obtained from the expectation values of the 
displacement operators 
\begin{equation} 
z_{L}^{c(s)} = \langle g|U_{L}^{c(s)}|g \rangle .  \label{zl} 
\end{equation} 
$|z_{L}^{c}|$ was first proposed by Resta and Sorella as an indicator of 
localization in extended systems.\cite{res2} By appealing to symmetry 
properties, Ortiz and one of the authors demonstrated that for 
translationally invariant interacting systems with a rational number 
$n/l$ of particles per unit cell the correct definition of the matrix 
element is $\langle g|(U_{L}^{c})^{l}|g \rangle$.\cite{znos} This 
definition was used to characterize metal-insulator and 
metal-superconducting transitions in one-dimensional lattice 
models.\cite{topo,znos,local} In the thermodynamic limit, $L \rightarrow 
\infty$, $|z_{L}^{c}|$ is equal to 1 for a periodic metallic system and 
to 0 for the insulating state. This result holds also for non-interacting 
disordered systems.\cite{local} For an intuitive understanding we note 
that in a metallic system with a well-defined Fermi surface, $U_{L}^{c}$ 
displaces the Fermi surface by a wave vector $2\pi / L$, and thus $\langle 
g|U_{L}^{c}|g \rangle = 0$. By contrast, in a non-interacting band 
insulator $U_{L}^{c}|g \rangle$ coincides with $|g \rangle$ and 
$z_{L}^{c} = 1$.  

The determination of $z_{L}^{c(s)}$ in the thermodynamic limit provides 
a means of calculating the Berry phases \cite{res2,znos,local,topo,epl} 
\begin{equation} 
\gamma_{c(s)} = \lim_{L \rightarrow \infty }{\rm Im} \ln z_{L}^{c(s)}. 
\label{berryz} 
\end{equation} 
For finite $L$, the right-hand side of Eq.~(\ref{berryz}) is actually 
equivalent to Eq.~(\ref{gamma}) if only a single point is used in 
the discretization of the flux. Thus it is not surprising that direct 
calculation of the Berry phase (or the MCEL) leads to a more precise 
extrapolated result than does Eq.~(\ref{berryz}).\cite{znos,topo} 
For comparison with these two types of analysis we have performed 
calculations of $z_{L}^{c(s)}$, as it emerges that these are important 
in determining the phase diagram of the AB$_{3}$ model at low values 
of $\Delta$. In addition, the displacement operators have a direct 
relationship with the bosonic fields introduced in Sec.~III: it was 
shown recently that $U_{L}^{c}$ and $U_{L}^{s}$ can be constructed 
as exponentials of the average charge and spin fields,\cite{poldi} 
\begin{equation} 
U_{L}^{\nu} = \exp (i \sqrt{8} \phi_{\nu }^{a}) \text{, } \phi_{\nu }^{a} 
 = \frac{1}{L} \int dx \, \phi_{\nu}(x).  \label{ucs} 
\end{equation} 
In the region of parameters such that the interaction $M_{c}$ in 
Eq.~(\ref{bhc}) is much greater than $m_{cs}$ [Eq.~(\ref{bhcs})], 
$\phi_{c}(x)$ becomes locked at the value which minimizes $M_{c} \cos 
(\sqrt{8} \phi_{c})$, leading to $\sqrt{8} \phi_{c}^{a} = \gamma_{c} = \pi$ 
(mod $2\pi $). If instead $m_{cs}$ is dominant, $\sqrt{8} \phi _{c}^{a} = 
\gamma_{c} = 0$. Similarly, by minimizing the potential for large positive 
$M_{s}^{\prime}$ one expects $\sqrt{8} \phi_{s}^{a} = \gamma_{s} = \pi$ 
(in fact $M_{s}^{\prime}$ renormalizes to zero when it is positive, but 
the result $\gamma_{s} = \pi$ is robust), while for negative $M_{s}^{\prime}$ 
one has instead $\sqrt{8} \phi _{s}^{a} = \gamma_{s}=0$. This is in agreement 
with the values obtained in the strong-coupling limit for the BI and the CI. 

In the intermediate FI phase the spin gap remains open, and thus the 
spin Berry phase $\gamma_{s}$ is zero, as in the BI. The situation with 
the charge Berry phase is more delicate: in the thermodynamic limit the 
inversion symmetry is broken spontaneously, leading to intermediate values 
of the polarization and also to fractional values of $\sqrt{8} \phi_{c}^{a}$, 
as has been shown in detail for $N = 2.$\cite{poldi} Both quantities are 
also related to the charge of the fractional excitations of the 
model.\cite{fab,poldi} However, for all finite systems, and in particular 
those of our numerical analysis, the ground state of the FI retains 
inversion symmetry and $\gamma_{c} = \pi$ as in the CI phase. 
 
In summary, as for the case $N = 2$,\cite{tor} the Berry phase vector  
${\bgamma}$ has the value $(0,0)$ in the BI phase, $(\pi,0)$ in the 
FI phase and $(\pi,\pi)$ in the CI phase. 
 
Finally, to investigate the possibility of a metallic phase near the 
charge transition, we have also calculated the Drude weight  
\begin{equation} 
D_{c} = \frac{L}{2} \frac{\partial^{2} E(L,\Phi)}{\partial \Phi^{2}}|_{\Phi 
= \Phi _{0}},  \label{drude} 
\end{equation} 
where $E(L,\Phi)$ is the energy of the ground state for a system of 
length $L$ with (anti-)periodic BCs in the presence of a flux $\Phi 
h c/(2\pi e)$ threading the ring. For the calculation of $U_{L}^{c(s)}$ 
and $D_{c}$ we have used BCs (or equivalently the choice of $\Phi_{0}$) 
corresponding to the ``closed-shell'' situation which gives the minimum 
in the energy as a function of flux. 

\begin{figure}[t!] 
\center \includegraphics[width =6.0cm,angle=-90]{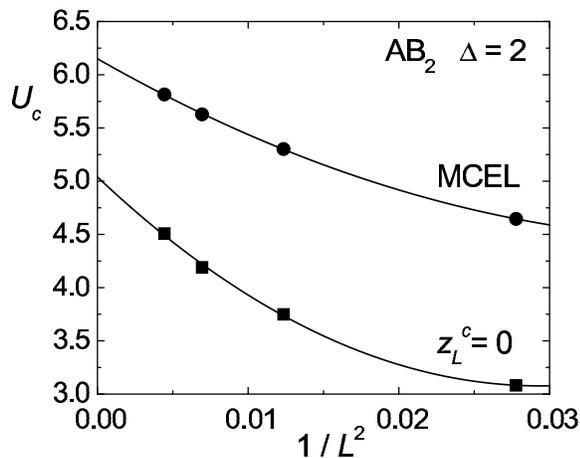} 
\caption{Critical value of on-site repulsion at the charge transition 
as a function of $1/L^2$, determined by the MCEL (circles) and by the 
condition $z_L ^c =0$ (squares) for the AB$_2$ chain with $\Delta = 2$.} 
\label{uc} 
\end{figure} 
 
\section{Results} 

To determine numerically the phase diagram of the AB$_{2}$ ($N = 3$) and 
AB$_{3}$ chains ($N = 4$) we have used both the MCEL, which corresponds to 
the jumps in charge and spin Berry phases (above), and the sign of 
$z_{L}^{c}$, both calculated by exact diagonalization with the Lanczos 
algorithm. The largest accessible system sizes are then $L = 15$ for 
$N = 3$ and $L = 16$ for $N = 4$, which are sufficient to allow a 
finite-size scaling analysis of limited accuracy. The two calculational 
methods should give the same result in the thermodynamic limit 
[Eq.~(\ref{berryz})], although as discussed in the previous section 
the former is usually more accurate.\cite{znos} However, we have found 
that for $\Delta \lesssim t$ the level crossings have large finite-size 
effects, particularly for $N = 4$, which limits the comparison of the 
two techniques. We have extrapolated the transition parameters using a 
quadratic polynomial in $1/L^{2}$, a dependence expected from conformal 
invariance and generally used in the MCEL.\cite{nom,nak,somma} 

\begin{figure}[t!] 
\center\includegraphics[width=6.0cm,angle=-90]{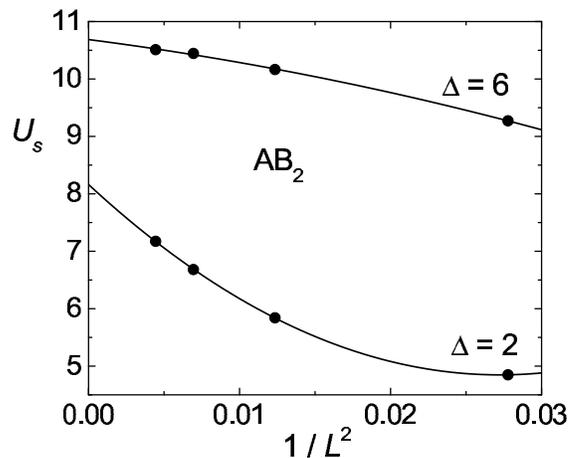} 
\caption{Critical value of on-site repulsion at the spin transition 
as a function of $1/L^2$, determined by the MCEL for the AB$_2$ chain 
with two values of $\Delta$.} 
\label{us} 
\end{figure}

Examples of the size-dependence of the resulting critical values at the charge 
($U_{c}$) and spin ($U_{s}$) transitions are shown in Figs.~\ref{uc} and 
\ref{us} respectively. The unit of energy is chosen as $t = 1$. For $U_{c}$ 
we have also tested an extrapolation with $1/L^{3}$ dependence,\cite{otsu} 
but the results were significantly inferior and displayed larger errors. 
For $\Delta > 2$ the error in the extrapolated value of $U_{c}$ obtained 
from the condition $z_{L}^{c} = 0$ is an order of magnitude larger than that 
obtained from the MCEL. Thus the MCEL may be taken to provide the more 
accurate results for $\Delta \ge 2$, as expected. For the majority of the 
parameter range these methods extrapolate acceptably well, and in a similar 
manner, towards a fixed value. However, for smaller values of $\Delta$, 
{\it i.e.}~in the weak-coupling regime, the quality of the fit deteriorates 
rapidly for both methods, but particularly for the MCEL. As discussed below, 
this method becomes unreliable for $N = 4$ and $\Delta < 1$. 

Figure \ref{us} shows the value of the spin transition ($U_s$) in the 
AB$_2$ chain for different system sizes, obtained using the MCEL. Again 
the data extrapolate in a satisfactory manner for larger values of 
$\Delta$, but a comparison of the two curves shows that the 
quality of the $1/L^{2}$ fit for the size dependence of $U_{s}$ also 
deteriorates with decreasing $\Delta$. A test of the MCEL based on 
conformal invariance and the renormalization group, and of the 
Kosterlitz-Thouless character of the spin transition, can be made by 
calculating the scaling dimensions $x_{\sigma ,1}$ of the singlet and 
$x_{\sigma ,2}$ of the triplet excitations.\cite{nom,nak} According to 
the predictions of conformal field theory, the excitation energies should 
scale according to  
\begin{equation} 
\Delta E_{\sigma ,i} \cong \frac{2 \pi v_{s}}{L} x_{\sigma ,i},  \label{ci} 
\end{equation} 
where $\Delta E_{\sigma ,1}$ ($\Delta E_{\sigma ,2}$) is the difference 
between the energy of the lowest singlet (triplet) state with BCs opposite 
to the closed-shell situation and the ground-state energy calculated with 
closed-shell BCs. The spin velocity $v_{s}$ is calculated from 
\begin{equation} 
v_{s} = \frac{E(1,2\pi /L) - E(0,0)}{2\pi /L},  \label{vs} 
\end{equation} 
where $E(S,K)$ is the lowest energy in the sector of total spin $S$ and 
total wave vector $K$. 

\begin{figure}[t!] 
\center\includegraphics[width=6.0cm,angle=-90]{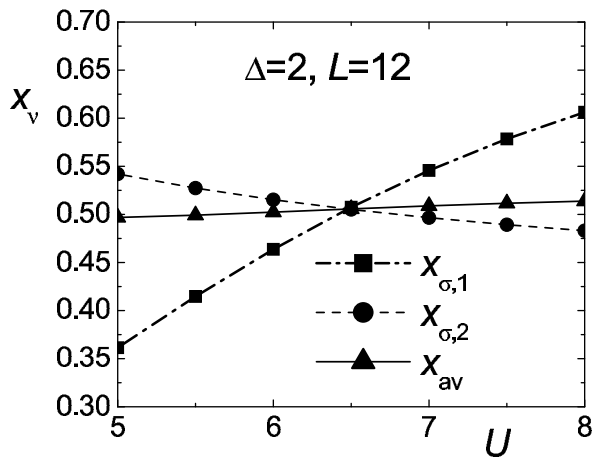} 
\caption{Scaling dimensions as a function of $U$ for the AB$_3$ chain 
with $L = 12$ and $\Delta = 2$.} 
\label{xnu} 
\end{figure} 

The values of $x_{\sigma ,1}$ and $x_{\sigma ,2}$ cross at $U_{s}$. Their 
average value $x_{av} = (x_{\sigma ,1} + 3 x_{\sigma ,2})/4$ should be 
equal to 1/2 near the crossing, and for $U > U_{s}$ where the spin gap 
is closed. Numerical results for the scaling dimensions are shown in 
Fig.~\ref{xnu}. Clearly $x_{av}$ is indeed close to 1/2, supporting the 
validity of the MCEL technique for the spin transition. For the reasons 
discussed in Sec.~IV a similar analysis at $U_c$ is less reliable.
Systems of larger size may be studied only by the density-matrix 
renormalization-group technique, which would offer a means of 
refining the quantitative aspects of the results below, as well as 
the possibility of investigating the critical behavior at the two
transitions.

\begin{figure}[t!] 
\phantom{.}
\vspace{-0.2cm}
\center\includegraphics[width=6.0cm,angle=-90]{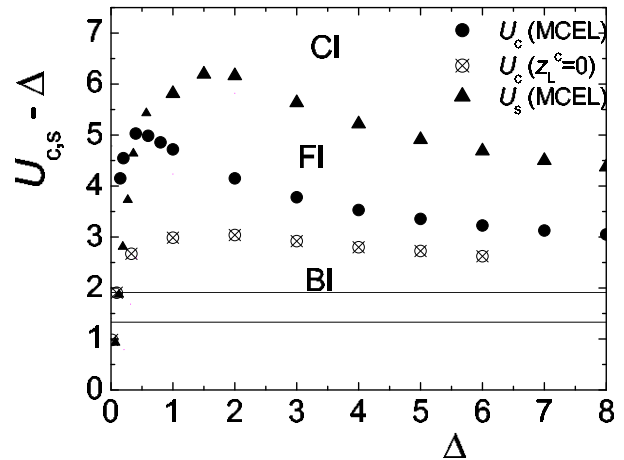} 
\caption{Phase diagram of the AB$_2$ chain.} 
\label{abb} 
\end{figure} 
 
\begin{figure}[b!] 
\center\includegraphics[width=5.8cm,angle=-90]{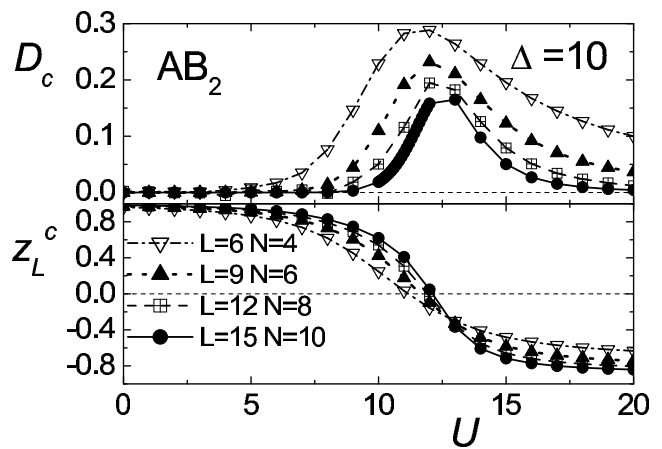} 
\caption{Drude weight and $z_L^c$ as functions of $U$ in the AB$_2$ 
chain with $\Delta = 10$ for a range of system sizes. The critical 
interaction strength obtained from the MCEL and by extraploating to 
the thermodynamic limit is $U_c = 12.94$ for this value of $\Delta$.} 
\label{dc2} 
\end{figure} 

Having demonstrated the application of the numerical methods discussed 
in Sec.~IV, we are now in a position to discuss the phase diagrams of 
AB$_{N-1}$ chains. In Fig.~\ref{abb} we show the extrapolated values 
of $U_{c} - \Delta$ and $U_{s} - \Delta$ as functions of $\Delta$ for 
the AB$_{2}$ system. The chain lengths used were $L = 6$, 9, 12, and 15. 
For $U_{c}$ we compare the results of the MCEL with those obtained from 
the change of sign of $z_{L}^{c}$ [cf.~Fig.~(\ref{uc})]. The difference 
between the two methods is significant, particularly for $\Delta \lesssim 
1$, where both methods have larger errors. For $\Delta < 0.75$, the points 
displayed for $U_{s}$, and for $U_{c}$ obtained from $z_{L}^{c}$,   
were calculated by fixing $U$ and extrapolating the critical value of  
$\Delta $. In comparison with the results obtained by fixing $\Delta $, 
this procedure shifts the curves to smaller values of $\Delta$, 
particularly for the MCEL. For the reasons stated above, at least for 
$\Delta \ge 2$, we take the MCEL results to be more accurate than those 
obtained from the change of sign of $z_{L}^{c}$. However, it is clear 
from Fig.~\ref{abb} that the two methods agree on a semiquantitative 
level. 

As expected from the strong-coupling analysis of Sec.~II, the differences  
$U_{c} - \Delta$ and $U_{s} - \Delta$ are of the order of the hopping $t$. 
For comparison, we have drawn two horizontal lines which correspond to the 
charge and spin transitions obtained with the MCEL for the AB chain (the 
conventional IHM\cite{tor}) for $t \ll U,\Delta$: $U_{c} - \Delta = 1.33t$ 
and $U_{s} - \Delta = 1.91t$. For the AB$_2$ chain $U_{c}$ is shifted to 
larger values, and also the difference $U_{s} - U_{c}$ becomes larger for 
$\Delta \gtrsim t$. Thus while there are no qualitative changes to the 
phase diagram, the widths of the BI regime and also of the intermediate 
FI phase increase on passing from AB to AB$_2$. This is in agreement 
with the conclusions of the bosonization treatment presented in Sec.~III. 
For $\Delta < t$, the numerical results show a tendency towards a negative 
difference $U_{s} - U_{c}$; this result is difficult to justify on physical 
grounds, such as those underlying the bosonization treatment, and is in all 
probability an artifact of finite-size effects, which are of the order of 
$|U_{s} - U_{c}|$ when $U_{s} - U_{c}$ is negative. 
 
Figure \ref{dc2} shows the dependence of the Drude weight $D_{c}$ and  
of the expectation value of the localization operator $z_{L}^{c}$ as 
functions of $U$ for $\Delta = 10t$ and for all system sizes studied. 
The evolution of these quantities with $L$ ($D_{c}$ decreasing and  
$|z_{L}^{c}|$ increasing) for $U \neq U_{c}$ indicates that the system 
is insulating over the entire range of parameters. However, because the 
system sizes are small one may not exclude metallic behavior at or near 
the point $U = U_{c}$. The vanishing value of $|z_{L}^{c}|$ at this point 
and the peak in $D_{c}$ are consistent with this possibility, but also 
simply with a larger localization length for $U \sim U_{c}$. From our 
theoretical and numerical analyses it is not possible to establish 
whether or not $D_{c}$ extrapolates to a finite value, which would 
indicate metallic behavior in the thermodynamic limit.
 
\begin{figure}[t!] 
\center\includegraphics[width=6.0cm,angle=-90]{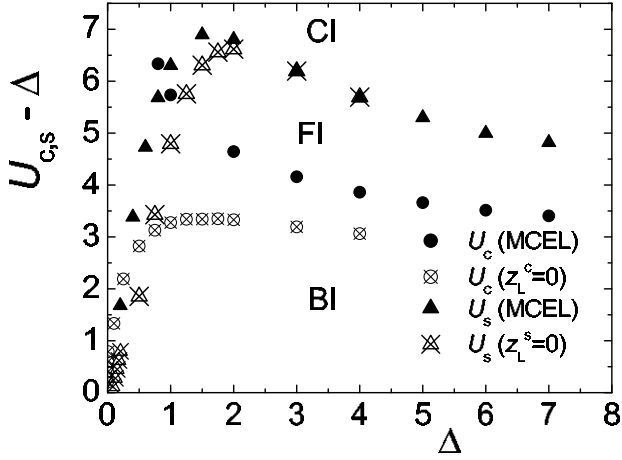} 
\caption{Phase diagram of the AB$_3$ chain.} 
\label{abbb} 
\end{figure} 
 
In Fig.~\ref{abbb} we show the extrapolated values of $U_{c} - \Delta$ 
and $U_{s} - \Delta$ as functions of $\Delta$ for the AB$_{3}$ chain. 
The system sizes used for this case were $L = 8$, 12, and 16, combined 
with a limited number of runs with $L = 20$. For $\Delta < 1$, all of 
the points displayed except those for $U_{c}$ obtained within the MCEL 
were calculated by fixing $U$ and extrapolating the critical value of 
$\Delta$. Surprisingly, $U_{c}$ obtained from the MCEL (or the jump in 
the charge Berry phase) appears to diverge as $\Delta \rightarrow 0$. 
To be specific, for $U = 50$ (not shown) the extrapolated value of 
$\Delta$ at the transition is $\Delta_{c} \simeq 0.6$, while for $L = 8$, 
$\Delta_{c} \simeq 3$; these results reflect the presence of very large 
finite-size effects. 

\begin{figure}[t!] 
\center \includegraphics[width=5.6cm,angle=-90]{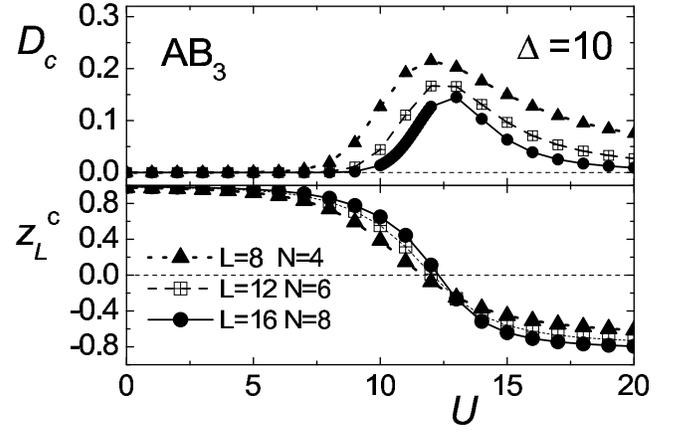} 
\caption{Drude weight and $z_L^c$ as functions of $U$ in the AB$_3$ 
chain with $\Delta = 10$ for a range of system sizes. The critical 
interaction strength obtained from the MCEL and by extraploating to 
the thermodynamic limit is $U_c = 13.20$ for this value of $\Delta$.} 
\label{dc3} 
\end{figure} 

This type of behavior is completely inconsistent with the strong-coupling 
results presented in Sec.~II, which show that for any $\Delta > 0$ and 
$U - \Delta \gg t$, the system is a CI with a charge distribution of the 
form represented schematically in Fig.~\ref{chd}(b). By contrast, the 
transition obtained from $z_{L}^{c}$ is consistent with the strong-coupling 
results. This fact, taken in combination with the very large finite-size 
effects of the MCEL for small $\Delta$, support the conclusion that the 
level crossing which is followed fails to describe the charge transition 
for $\Delta \lesssim 1$. This difficulty may be related to the loss of 
translational symmetry in comparison with the Hubbard model. In fact, 
in the example of the extended Hubbard model with correlated hopping, 
where a similar charge transition occurs,\cite{tor,bos,topo,nak} the 
quantum numbers for the appropriate crossing energy levels can be 
identified from a weak-coupling analysis and correspond to wave vector 
$K = \pi/a_0$.\cite{nak} When the one-site translational symmetry is lost, 
this wave vector is mixed with others and the Lanczos method, which 
captures the lowest eigenstate for the corresponding wave vector in 
the reduced Brillouin zone, no longer follows necessarily the correct 
excitation in the weakly coupled case. This problem arises also in the 
extended IHM.\cite{tor} 

We proceed by disregarding the points obtained for the charge transition 
with the MCEL for $\Delta \lesssim 1$, and in this regime adopt the 
values obtained from $z_L^c$. The phase diagram of the AB$_{3}$ chain 
(Fig.~\ref{abbb}) is very similar to that of AB$_{2}$. Only at the 
quantitative level is there a moderate shift of $U_{c}$ and $U_{s}$ to 
still larger values and a small increase in the width of the region 
occupied by the FI phase. In Fig.~\ref{abbb} we show also the values 
of $U_s$ for the spin transition obtained from the condition $z_{L}^{s}
 = 0$. These are in very close coincidence with the results obtained from 
the MCEL for $\Delta \ge 2$; for small $\Delta$ both methods are 
again plagued by large finite-size effects, but the discrepancy between 
the extrapolated values is significantly smaller than in the case of the 
charge transition. 
 
Finally, Fig.~\ref{dc3} displays the dependence of $D_{c}$ and $z_{L}^{c}$ 
on $U$ in an AB$_3$ chain with $\Delta = 10t$ for several system sizes. 
The results are qualitatively and quantitatively very similar to those 
obtained for the AB$_2$ system (Fig.~\ref{dc2}). At $U > U_c$ some 
indications of a smaller gap and longer localization length might have 
been expected as the size of the unit cell $N$ increases, as suggested 
by the results of Sec.~II, but in this respect no significant differences 
are found between the AB$_{2}$ and AB$_{3}$ chains investigated here. 
 
\section{Summary and discussion} 

We have performed a systematic analysis of the generalized ionic Hubbard 
model (IHM) in one dimension. Motivated by the complex behavior emerging 
in the AB chain, where the commensurate filling ensures a dominant role 
for Umklapp scattering terms, we have investigated the phase diagrams for 
AB$_{N-1}$ chains with unit cell size $N$ and filling $1/N$. We have 
employed analytical considerations in both weak- and strong-coupling 
limits, and sophisticated numerical techniques based on Lanczos exact 
diagonalization, applying these specifically to the cases $N = 3$ and 
$N = 4$. 

Despite the absence of direct Umklapp terms away from $N = 2$, we find 
that the qualitative features of the AB$_{N-1}$ phase diagram are 
essentially identical to those of the conventional IHM. The generalized 
model always possesses three phases as a function increasing values of 
$U$: a band insulator (BI) for small $U/\Delta$, a correlated insulator 
(CI) for large $U/\Delta$ which is a form of Mott insulator, and between 
these an intermediate ferroelectric insulator (FI) with spontaneous 
breaking of inversion symmetry and fractional excitations. The generic 
nature of the phase diagram is explained within a bosonization analysis 
by the emergence of higher-order interaction terms related to the 
charge-density modulation caused by the ionic term. 

The general similarity of all the models in this class masks some 
important differences. Qualitatively, the CI phase for large $U$ is 
not strictly a Mott insulator because the number of particles is less 
than the number of sites. As a quantitative consequence, the localization 
length is therefore larger than one lattice parameter and the gap is much 
smaller than in the conventional ($N = 2$) IHM. In the latter, for large 
$U$ the gap is of order $U - \Delta$, whereas for $N > 2$ it is given 
by Eq.~(\ref{gapsc}) and vanishes for $\Delta =0$. More generally, the 
relatively weaker Umklapp terms result in larger values of $U$ being 
required to compete with the ionic term $\Delta$, which favors the BI 
phase, and thus the FI and CI phase boundaries are pushed to larger $U$ 
with increasing $N$; both BI and FI regimes are broadened.  

We have implemented the numerical calculation of a number of 
characteristic quantities which provide valuable insight into 
the properties of the ground and excited states. For the two 
phase boundaries we have employed the method of crossing excitation 
levels (MCEL), which is equivalent to following discontinuous steps in 
the charge and spin Berry phases, quantities which signal the presence 
of a transition even on a small system. This calculation is compared 
with the results obtained from the vanishing of the expectation value 
of the displacement operators, which is found to be a less accurate
but sometimes more reliable technique. We have also computed the 
Drude weight with a view to testing the possible metallic nature 
of the system at the phase boundaries between the insulating phases.

While the numerical analysis serves as a useful test of these 
techniques, and allows us to make certain powerful qualitative 
statements, the results are also subject to strong finite-size effects.
In particular, the MCEL has significantly larger finite-size effects 
for $N > 2$ than at $N = 2$, due probably to the much smaller gap in 
the CI phase and the relatively lower translational symmetry. Indeed, 
at the charge transition for $\Delta < t$ the phase of $z_{L}^{c}$ 
gives more reliable information. However, even with the restricted 
system sizes accessible in our study, the majority of the results 
obtained in Sec.~V can be said to be of semi-quantitative accuracy. 
We comment that further studies on systems of larger size are possible 
using the density-matrix renormalization-group technique: with the 
aid of the quantities we have defined and analyzed in Secs.~IV and V, 
such calculations might confirm our findings with greater accuracy, 
establish more precisely the boundary of the charge transition, and 
possibly also permit the analysis of critical behavior at the two
transitions.

In the same way that the AB chain has for many years been of interest 
in the context of quasi-one-dimensional organic $MX$ complexes, our 
considerations concerning the AB$_2$ chain are expected to be relevant 
for the properties of $MMX$ chains. These are halogen-bridged binuclear 
metal chains such as R$_4$[Pt$_2$(P$_2$O$_5$H$_2$)$_4$X.$n$H$_2$O, on 
which experimental studies would be welcomed. We comment that the 
greater extent in parameter space of the FI phase for $N > 2$ systems 
may even be important for technological applications. 
Finally, with regard to the possible metallic properties of AB$_{N-1}$ 
systems, we have found that despite being at fillings $1/N$ away from 
the half-filled situation, these remain insulating for all parameter 
regimes other than $\Delta = 0$, as a consequence of the commensurate 
nature of their electron number. Only at the charge transition do we 
obtain evidence from the Drude weight and from the charge localization 
operator of a possible metallic point in the phase diagram. 

\section*{Acknowledgments} 
 
This work was sponsored by PICT 03-12742 of ANPCyT. M.E.T. and A.A.A. are
partially supported by CONICET. G.I.J. would like to acknowledge the 
generous support and hospitality of the MPI-PKS Dresden, where part of 
this work was performed.

\end{document}